\documentstyle[12pt,world_sci]{article}

\title{On Collinearization of Quarks \\
in Quark-Gluon Decays of Heavy Orthoquarkonia}

\author{A.Ya.Parkhomenko and A.D.Smirnov \\
{\small\it Department of Theoretical Physics, Yaroslavl State University,}\\
{\small\it Sovietskaya 14, 150000 Yaroslavl, Russia}}

\date{}

\begin{document}

\begin{flushright}
{\normalsize Yaroslavl State University\\
             Preprint YARU-HE-94/07\\
             hep-ph/9409449} \\[3cm]
\end{flushright}

\maketitle

\begin{abstract}

The decay of heavy orthoquarkonium into quark-antiquark pair and two
gluons is considered. The differential probability of the decay
in the tree approximation is calculated and the final quark mass influence
on the quark and gluon distribution functions is studied. It is shown that
the collinear strengthening of the bottomonium decay probability takes place
at the $u \bar u$-, $d \bar d$- and $s \bar s$- pair production
and is absent at the $c \bar c$- production.

\end{abstract}

\vglue 3cm

\begin{center}
{\it Talk given at the VIII International Seminar "Quarks-94",\\ Vladimir,
Russia, May 11-18, 1994}
\end{center}

\newpage

\indent The decays of heavy quarkonia give  the  useful  information
on the dynamics of quarks and gluons and  on  the  processes  of  jet
production of hadrons. In  particular  the  many-particle  decays
such as $^{1}S_{0} \rightarrow 3g, \; q \bar q g$  and $^{3}S_{1}
\rightarrow 4g, \; q \bar q g g$ are of great interest as
the processes  giving the immediate information on  $q \bar q g$- and $3g$-
interactions. The nonabelian nature of  ggg-interaction  manifest
itself, for example, in the distribution on invariant  masses  of
two particles~\cite{KSWZ,St}, as the acomplanarity  of four particle
decays~\cite{MN}, as the collinearization of the gluons~\cite{Sm}
and as some other effects. It is worth noting that the four particle decays
of heavy  orthoquarkonia  are  more  available  for  experimental
investigations because of  rather  great  probability  of  direct
production  of orthoquarkonia  in $e^{+}e^{-}$- and $p\bar{p}$--
collisions. By this reason the detail theoretical analysis of  these
decays is interesting on purpose to find the optimal conditions
for their experimental observation.

In this work we consider the decay of heavy  orthoquarkonium
with  spin-parity $J^{PC}=1^{- \, -}$  into  quark-antiquark  pair  and  two
gluons.  The  amplitude  and  differential  probability  of  this
process  in  tree  approximation  and  some  angle   and   energy
distributions of quarks and gluons are obtained and discussed.

The process $^{3}S_{1}(Q\bar{Q}) \rightarrow q \bar q g g$ is described
in tree approximation by six diagrams shown on Fig.~1. The amplitude of
this process in the initial static quarks limit in the rest frame
of quarkonium can be presented in the form:

\begin{eqnarray}
M^{\alpha\beta}_{ab} \big ( ^3S_1 (Q\bar{Q}) \rightarrow q{\bar q}gg \big ) =
{ d_{abc} (t_c)_{\alpha\beta} \over 4 \sqrt N_c}
{g^4_{st} \psi(0) \over 4 \sqrt{\epsilon_1 \epsilon_2 \omega_1 \omega_2}}
{\big ( w^T_{\bar Q} \sigma_2 \vec{\sigma} w_Q \big ) \over 4 m^2 p^2}
{\vec V},
\end{eqnarray}

\begin{figure}[b]
\unitlength=1.00mm
\begin{picture}(155,50)
\put(30,0){\parbox[b]{0.55\hsize}{\caption{Diagrams of the decay
            $^3S_1 (Q \bar Q) \to q \bar q g g$ in the tree approximation.}}}
\end{picture}
\end{figure}

\noindent where $d_{abc}$ ($a,b,c = 1,2,\ldots,N^2_c-1$) are symmetrical
constants of $SU(N_{c})$, $t_{c}$ are generators of $SU(N_{c})$ group
normalized as $Sp(t_{a}t_{b}) =
\delta_{ab}/2$, $\psi(0)$ is the nonrelativistic wave function of quarkonium
in  the
coordinate space, $w_{Q}$ and $w_{\bar Q}$ are two-component spinors of initial
quark and antiquark, $p = p_{1}+ p_{2}$, $p_{1}$ and $p_{2}$ are 4-momenta
of final  quark  and antiquark, $m$ is a mass of the initial quark,
$g_{st}$ is a color charge,
$\vec \sigma = (\sigma _{1}, \sigma _{2}, \sigma _{3})$ are Pauli
matrices.

The vector function $\vec V$ in (1) depends on momenta and polarizations
of final quarks and gluons and in the three-transversal gauge takes
the rather simple form:

\begin{eqnarray}
\vec V & = &
\big ( \vec e^+_1 \vec e^+_2 \big ) \vec j^- +
\big ( \vec e^-_1 \vec e^-_2 \big ) \vec j^+ +
\big ( \vec e^+_1 \vec j^+ \big ) \vec e^-_2 +
\big ( \vec e^-_1 \vec j^- \big ) \vec e^+_2 \nonumber \\ & + &
\big ( \vec e^+_2 \vec j^+ \big ) \vec e^-_1 +
\big ( \vec e^-_2 \vec j^- \big ) \vec e^+_1 +
4 \Big \{ \Big [ 1 - \frac{2mp_0}{(kp)} \Big ] \vec j +
\frac{2m\vec{p}}{(kp)} j_0\Big \} \\ & + &
i j_0 \Big [ 1 - \frac{4m^2}{(kp)} \Big ] \Big \{ \big [ \vec e^+_1,
\vec e^-_2 \big ] - \big [ \vec e^-_2, \vec e^+_1 \big ] \Big \}, \nonumber
\end{eqnarray}

\noindent where $\vec e^{\pm}_{i} = \vec e_{i} \pm i \big [ \vec n_{i},
\vec e_{i} \big ]$, $\vec n_{i} = \vec k_{i}/\omega_{i}$, $e_{i} = \big \{ 0,
\vec e_{i} \big \}$, $k_i = \big \{ \omega_i, \vec k_i \big \}$
are 4-vectors  of polarization  and momentum  of $i$-th gluon  ($i=1,2$),
$\big ( \vec n_i \vec e_i \big ) = 0$, $\vec j^{\pm} = \vec j \pm
2 m i \big [ \vec j, \vec p \big ] / (kp)$, $k = k_{1}+ k_{2}$,
$j  = \big \{ j_0,\vec j \big \} = \bar u (p_1) \gamma u(-p_{2})$ is a
four current of final $q\bar{q}$-pair, $u(p_{1})$ and $u(-p_{2})$ are Dirac
spinors  of final quark and antiquark, $\gamma  = \big \{ \gamma_0,
\vec \gamma \big \}$ are Dirac matrices.

The  corresponding  to~(1) and~(2)   differential   probability
averaged over three spin states of  initial  orthoquarkonium  and
summed over polarizations and colors of final  particles  may  be
presented in the form:

\begin{eqnarray}
dW = {F_c \over 6 \pi^4} {\alpha^4_s \mid \psi(0) \mid^2 Q_{\mu \nu}
G_{\mu \nu} \over \big [ p^2 (kp) (Pk_1) (Pk_2) \big ]^2}
\delta (P-p-k) {d\vec{p_1}d\vec{p_2}d\vec{k_1}d\vec{k_2} \over \epsilon_1
\epsilon_2 \omega_1 \omega_2},
\end{eqnarray}
\begin{eqnarray}
F_{c} = {\big ( N^{2}_{c} - 4 \big ) \big ( N^{2}_{c} - 1 \big ) \over
32 N^{2}_{c}}, \nonumber
\end{eqnarray}

\noindent where $F_c$ is the color factor of $SU(N_{c})$- group,
for $SU(3)$ $F_{c}= 5/36$,
$P = \big \{ 2m,\vec 0 \big \}$ is the four momentum of
quarkonium in the rest frame, $\alpha_s = g^{2}_{st} / (4\pi)$ is the strong
coupling constant, $Q_{\mu \nu}$, $G_{\mu \nu}$ are symmetrical tensor
functions of momenta of quarks and gluons correspondingly. The expressions
for $Q_{\mu \nu}$ and $G_{\mu \nu}$ obtained by us have the structure
such as that of the tensors $L_{\mu \nu}$, $H_{\mu \nu}$ of Ref.~\citenum{CCH}.

Integrating~(3) over quark and antiquark momenta we obtain the probability
distribution in the energies $x_{i} = \omega_{i}/m$ of the gluons and
angle $\vartheta _{g}$ between their momenta in the form:

\begin{eqnarray}
{dW \over dx_1 dx_2 d\cos\vartheta_g} = F_c
{\alpha_s^4 \mid \psi(0) \mid^2 \over 36 \pi m^2} {F_g \over \eta_g \xi^2_g}
\Big ( 1 + {2 \mu^2 \over \eta_g} \Big ) \sqrt{1- {4 \mu^2 \over \eta_g}} ,
\end{eqnarray}
\begin{eqnarray}
F_g & = & 8 x_1 x_2 \big [ 12 \big ( 1 + \cos^2 \vartheta_g \big )
- 8 \big ( x_1 + x_2 \big ) \big ( 1 - \cos \vartheta_g + \cos^2
\vartheta_g \big ) \nonumber \\
& + & 4 \big ( 1- \cos \vartheta_g \big ) \big [ 2 \big ( x_1 + x_2 \big )^2
- x_1 x_2 \big ( 1 - \cos \vartheta_g - \cos^2 \vartheta_g \big ) \big ]
\nonumber \\
& - & 8 x_1 x_2 \big ( x_1 + x_2 \big ) \big ( 1 - \cos^2 \vartheta_g \big )
+ x^2_1 x^2_2 \big ( 1- \cos \vartheta_g \big )^3 \big ( 3 - \cos \vartheta_g
\big ) \big ], \nonumber \\
\eta_g & = & (P-k)^2 / m^2 = 4 \big ( 1 - x_1 - x_2 \big ) + 2 x_1 x_2
\big ( 1 - \cos \vartheta_g \big ), \nonumber \\
\xi_g & = & \big ( (Pk) - k^2 \big ) / m^2 = 2 \big ( x_1 + x_2 \big )
- 2 x_1 x_2 \big ( 1 - \cos \vartheta_g \big ), \nonumber
\end{eqnarray}

\noindent where $\mu = m_q / m$ is mass ratio of final and initial quarks.

Similarly integrating~(3) over momenta of gluons we obtain the probability
distribution in the energies $y_i = \epsilon_i / m$ of quarks and angle
$\vartheta_q$ between their momenta in the form:

\begin{eqnarray}
{dW \over dy_1 dy_2 d\cos\vartheta_q} = F_c {\alpha_s^4 \vert \psi(0)
\vert^2 \over 36 \pi  m^2} {F_q \over \eta^2_q \xi^2_q \big
( 2 - y_1 - y_2 \big ) v^4} ,
\end{eqnarray}
\begin{eqnarray}
\eta_q & = & p^2 / m^2 = 2 \Big [\mu^2 + y_1 y_2 - \cos \vartheta_g
\sqrt{\big ( y_1^2 - \mu^2 \big ) \big ( y_2^2 - \mu^2 \big )} \Big ],
\nonumber \\
\xi_q & = & \big ( (Pp) - p^2 \big ) / m^2 = 2 \big ( y_1 + y_2 \big ) -
\eta_q, \nonumber \\
v & = & \sqrt{1 - \frac{4 m^2 k^2}{(Pk)^2}} =
\frac{\sqrt{( y_1 + y_2 )^2 - \eta_q}}{2 - y_1 - y_2}, \nonumber
\end{eqnarray}

\noindent where $v$ is the center-of-mass velocity of the gluon pair in the
rest frame of quarkonium,

\begin{eqnarray}
F_q = P_1 (y_1, y_2, v^2) + { 1 - v^2 \over v} \ln \Big\vert {1+v \over 1-v}
\Big\vert P_2 (y_1, y_2, v^2),
\end{eqnarray}

\begin{figure}
\unitlength=1.00mm
\begin{picture}(150,95)
%
%
%
\put(5,22){\parbox[t]{0.40\hsize}{\caption{Angle distributions of
gluons at $x_1 = x_2 = x$ and quarks
at $y_1 = y_2 = y$ in the decay $^3 S_1 (b \bar b) \to s \bar s g g$:
a) $x = 0.4$ or $y = 0.4$, b) $x = 0.6$ or $y = 0.6$,
c)~$x = 0.8$ or $y = 0.8$.}}}
\put(80,-3){\parbox[b]{0.41\hsize}{\caption{Angle distributions of
gluons at $x_1 = x_2 = x$ and quarks
at $y_1 = y_2 = y$ in the decay $^3 S_1 (b \bar b) \to c \bar c g g$:
a) $x = 0.4$ or $y = 0.4$, b) $x = 0.6$ or $y = 0.6$,
c)~$y = 0.8$.}}}
\end{picture}
\end{figure}

\noindent $P_1$ and $P_2$ are some polynomials of degree six in $v^2$, which
have a rather complicated form and we don't adduce them here.

It's convenient further to  use  the  distributions~(4) and~(5)
normalized   as

\begin{eqnarray}
f_g & = & \frac{1}{\alpha_s W_{3g}} \frac{dW}{dx_1 dx_2 d\cos\vartheta_g},
\nonumber \\
f_q & = & \frac{1}{\alpha_s W_{3g}} \frac{dW}{dy_1 dy_2 d\cos\vartheta_q}, \\
W_{3g} & = & 2 F_c \frac{16 \big ( \pi^2 - 9 \big )}{9}
\frac{\alpha^3_s \mid \psi(0) \mid^2}{m^2}, \nonumber
\end{eqnarray}

\noindent where $W_{3g}$ is the three-gluonic decay probability of
orthoquarkonium. The gluon (quark) distributions $f_g$ ($f_q$) as the
functions of angle $\vartheta_g$ ($\vartheta_q$) between the momenta of the
gluons (quarks) at equal energies $x_1 = x_2 = x$ ($y_1 = y_2 = y$)
are presented on Fig.~2 and~3 for bottomonium decays $^3S_1 (b\bar{b})
\rightarrow s \bar s g g$ and $^3S_1 (b\bar{b}) \rightarrow c \bar c g g$
correspondingly. These distributions exhibit two interesting peculiarities.

First, the analysis of the quark distribution $f_q$ shows that
the probability of the production of light $q \bar q$- pair and two
gluons can significantly increase as  the  angle  between  quarks
decreases. This collinear  effect  has  a  simple  nature  -- the
decrease of the denominator in  the  propagator  of  the  virtual
gluon as the angle between  light quarks decreases. As  shown  in
Fig.~2,~3 this effect exhibits itself in the bottomonium decay with
production of $s \bar s$- pair (it's more exhibited in the decays with
production of $u \bar u$-, $d \bar d$- pairs)
but it's absent in the decay with
production of  heavy $c \bar c$- pair. It means, that in the decays
$^3S_1 (b\bar{b}) \rightarrow q \bar q g g$ of bottomonium the production of
light $u \bar u$-, $d \bar d$- and $s \bar s$- pairs at small angle
between quark
and antiquark is more probable than that of $c \bar c$- pair. The
corresponding 4-jet events can be possibly available for experimental
observation by appropriate angle resolution.

Secondly, the production of hard $q \bar q$- pair  with  large  angle
between $q$- and $\bar q$- quarks accompanied by production  of  two  soft
gluons is more probable too (see quark curve c) on Fig.~2,~3).

The angle distributions of the quarks  and  gluons  obtained
and discussed by us may be useful for the  experimental  searches
and investigations of the four-jet  events  from  the  decays  of
heavy quarkonia.

\vspace{\baselineskip}

\noindent {\bf Acknowledgment}

\vspace{0.5\baselineskip}

\noindent This work was supported by the Russian Foundation for
Fundamental Research (Grant No. 93-02-14414).

\end{document}